\documentclass[aps,showpacs,twocolumn]{revtex4}
\usepackage{bm}
\usepackage{amsmath}
\usepackage{amssymb}
\usepackage{graphicx}
\usepackage{multirow}

\usepackage{titlesec}
\usepackage{lipsum}

\titleformat{\section}
  {\normalfont\bfseries}{\thesection}{1em}{}

\begin{document}

\def\pr{\prime}
\def\be{\begin{equation}}
\def\en#1{\label{#1}\end{equation}}
\def\d{\dagger}
\def\bar#1{\overline #1}
\def\K{\mathcal{K}}
\def\S{\mathcal{S}}
\def\D{\mathcal{D}}
\def\W{\mathcal{W}}
\def\E{\mathcal{E}}
\newcommand{\per}{\mathrm{per}}

\newcommand{\rd}{\mathrm{d}}
\newcommand{\vare}{\varepsilon }

\title{ Universality of generalized bunching and     efficient  assessment   of   Boson Sampling   }

\author{V. S. Shchesnovich}

\affiliation{Centro de Ci\^encias Naturais e Humanas, Universidade Federal do
ABC, Santo Andr\'e,  SP, 09210-170 Brazil }

\begin{abstract}
It is found that identical  bosons (fermions)  show    generalized    bunching   (antibunching)  property  in    linear  networks:   The absolute   maximum  (minimum)  of  probability that all $N$ input particles are detected   in a  subset of  $\K$ output modes of any nontrivial linear $M$-mode network is attained \textit{only} by completely indistinguishable  bosons (fermions).   For fermions $\K$ is arbitrary, for bosons it is either ($i$) arbitrary for  only classically correlated bosons or ($ii$) satisfies $\K\ge N$ (or $\K=1$) for arbitrary input states of $N$ particles.      The generalized bunching allows to certify in a \textit{polynomial} in $N$ number of runs   that a physical device realizing  Boson Sampling with  \textit{an arbitrary} network operates in the regime of full   quantum coherence compatible \textit{only} with completely indistinguishable  bosons. The protocol  needs \textit{only polynomial} classical computations  for the standard BosonSampling, whereas  an \textit{analytic formula} is available   for  the Scattershot version. \end{abstract}

\maketitle

 \textit{Introduction.--} Optical networks with photons   have become a growing   research field with potential application in   quantum computing \cite{KLM,AA,MVS}.   Boson Sampling (BS) idea \cite{AA},  a non-universal but  near-future  feasible  device aiming  at the Extended Church-Turing thesis (ECT),  followed  by   spectacular     experiments with optical networks  of  growing  size and number of photons  \cite{E1,E2,E3,E4,E5,E6,E7}, can  be a  way for benchmark demonstration of quantum supremacy.    A BS device with a random but known  $M$-mode network,  $N\sim 30$   single photons at known   input modes   for  $M\gg N^2$ would achieve this ultimate goal \cite{AA}.  However,  the very computational complexity of BS \cite{Valiant,A}  requires an exponential in $N$ number of runs of such a device    and computation of an exponential number (at least $O(N^N)$) of classically hard permanents to prove BS  by comparing an output distribution with theoretical probabilities.   Quantum supremacy demonstration thus faces  a big  challenge of maintaining  the  $N$th order quantum coherence in a BS device with $N\sim 30$    for an exponential number of runs.  Though a distribution claimed to simulate  BS, as the uniform one  \cite{Gogol},  can be exposed  with only  polynomial number of runs  of a device \cite{AA2} (see also Ref. \cite{E6}), finding such a  protocol for  a given sampler, e.g.,   the one with distinguishable particles,  is a hard  open problem.

The proof of BS   being  exponential  both in the number of runs and computations does not prevent efficient   verification of the  very  source of  quantum supremacy of BS, i.e.,  the full $N$th-order quantum coherence in  a device with $N$ bosons,  where unwanted distinguishability, photon losses, and  higher photon numbers are the leading  adversary factors in optical setups    \cite{E1,E2,E3,E4,E5,E6,E7}.  Such an assessment  of BS may require only a polynomial number of runs  and polynomial  classical computations.     Zero-transmission laws in the Fourier   network \cite{ZlawBS} and statistical benchmarking \cite{StatB}  probed this path, but  an assessment protocol applicable to \textit{an arbitrary network} is an open problem. The  test of Ref. \cite{ZlawBS}  does verify the  $N$th order coherence but only in   a  single Fourier network (not posing a threat to the ECT), whereas    the  statistical method of Ref.   \cite{StatB}  can only assess  the  second-order  coherence.  Already with relatively  small  networks   experimentalists have to   resort to either  Bayesian methods  or circumstantial evidence \cite{E5,E6,E7}.

\begin{figure}[ht]
\begin{center}
\includegraphics[width=0.425\textwidth,height=0.25\textwidth]{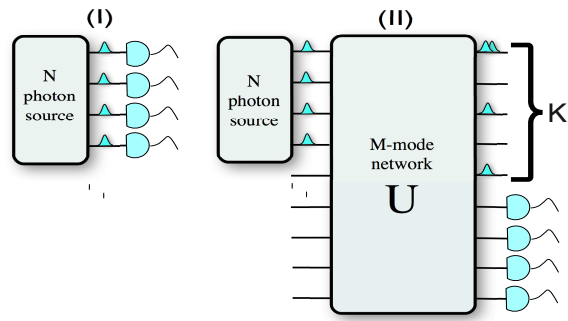}
\caption{(Color online) Assessment  protocol for a BS device.  Step (I):  the  source is   checked  for   output with only one particle per mode. Step (II), the   statistics  of all $N$ input particles to land in   $\K$ output modes of a $M$-mode network $U$ is gathered by   on-off type  detectors in complementary $L = M-\K$ modes:  The  maximum probability   is attained \textit{only} by completely indistinguishable bosons (and is close to $1$ for  $NL \ll M$).  }\label{F0}
\end{center}
\end{figure}

\textit{Generalized bunching  and efficient assessment of BS.--}        We call  a protocol that uses  only  polynomial classical computations  to certify in a   polynomial number of runs that a BS device operates at the full quantum  $N$th-order  coherence  compatible \textit{only} with completely indistinguishable bosons an efficient assessment protocol for BS.   Efficiency  \textit{requires} an \textit{independent} network certification:  An efficient assessment of BS  not using certification of a network matrix in a regime different from BS allows  for  loopholes, due to a combined effect of    network and distinguishability errors  (see the Appendix).   This conclusion agrees  with the results of  Refs. \cite{Gogol,AA2}.  An assessment protocol for a BS device with an \textit{arbitrary} network  can only be based on  a network-independent (universal)  \textit{absolute  maximum or minimum} of some probability attained \textit{only} by completely indistinguishable bosons~\footnote{E.g., average $N$-fold detection probability of  indistinguishable bosons is less than that of distinguishable particles \cite{E5}, whereas a  BS assessment protocol  needs an absolute minimum over \textit{all} inputs with partially distinguishable  particles \textit{in any non-trivial network}.}. The discovered  generalized boson  bunching is such a property:   The  absolute  maximum of  probability  of detecting  \textit{all} $N$ input particles  in   $\K $ output modes of an arbitrary nontrivial  $M$-mode  linear network  is attained \textit{only}  by    completely indistinguishable bosons    ($i$)  over arbitrary input states of particles  for $\K\ge N$ (and $\K=1$)  or  ($ii$) for arbitrary $1\le \K\le M-1$   over   only classically-correlated  bosons.   The minimum probability is attained by    completely indistinguishable fermions, with that of distinguishable particles lying in the middle between the two. The   generalized bunching/antibunching   is found by  exploiting  the discovered      equivalence  between  probability  of detecting all input  particles in  a subset of  output modes of  a linear   network  and an   eigenvalue problem for  positive semi-definite (p.s.d.)  Hermitian  matrix (Eq. (\ref{pN}) below).

The generalized bunching can be viewed as an  $N$-particle  generalization  of   the Hong-Ou-Madel (HOM) effect \cite{HOM}  to arbitrary $M$-mode networks.  Known generalizations of the HOM effect   to many-particle multi-mode setups  are reported for  special   Bell  multiport  networks  \cite{LB,Ou2008,MPBF}, whereas  bunching to a single mode  of a  random network \cite{GenBunch} has   probability  decaying   as  $\sim e^{-N}$ for     $M\ge N^2$  \cite{AK}.     A similar effect   is boson  tendency to form clouds in output modes in \textit{some  networks}  \cite{E5}.  \textit{ In contrast,   the generalized bunching   can be   observed in a polynomial in $N$ number of experimental runs in an  arbitrary (nontrivial) quantum network.}  Though  in the dilute  limit, $M\ge N^2$, the generalized  bunching/antibunching effect  wanes  as $N\to\infty$, it nevertheless  allows for   an efficient    assessment    protocol  of  a BS device, sketched in Fig.~\ref{F0},  requiring, besides only a polynomial number of runs of a device,  polynomial classical computations (see below). Moreover,  analytical results are derived for the Scattershot version of BS  \cite{SCBS}.

\textit{Description of partially distinguishable identical particles in a linear network.--}
 We  consider an $M$-mode  linear quantum network with $N$ single identical particles in an arbitrary internal   state at  input modes $k_1,\ldots,k_N$  (ordered products of operators  are assumed  in case of fermions) 
 \be
 \rho = \sum_{i}q_i|\Psi_i\rangle\langle\Psi_i|,\quad 
|\Psi_i \rangle = \sum_{\mathbf{j}}C^{(i)}_{\mathbf{j}}\prod_{\alpha=1}^N a^\dag_{k_\alpha,j_\alpha}|0\rangle,
\en{Psi}
where    $q_i\ge 0$, $\sum_i q_i =1$,  $\mathbf{j} = (j_1,\ldots,j_N)$, $\sum_{\mathbf{j}} |C^{(i)}_{\mathbf{j}}|^2 =1$, and   a mode operator  $ a^\dag_{k,j}$  (and below  $b^\dag_{k,j}$)   creates a particle in an input (output) mode $k$ and an internal basis state $|j\rangle \in\mathcal{H}$ (e.g., a basis  function  of  spectral shape of a photon).  A unitary network with matrix  $U$, Fig. \ref{F0}(b),   relates  the input $a_{k,j}$ and output $b_{l,j}$ modes: $a^\dag_{k,j} = \sum_{l=1}^{M} U_{k,l}b^\dag_{l,j}$.  The internal state of identical particles, defined as  
\be
 \rho^{(int)}  = \sum_i q_i |\psi_i\rangle\langle\psi_i|,\quad |\psi_i\rangle \equiv \sum_{\mathbf{j}} C^{(i)}_{\mathbf{j}}  \prod_{\alpha=1}^N{\!}^{\otimes}|j_\alpha\rangle,
\en{Int}
governs their behavior in a linear network. Symmetry properties of $\rho^{(int)}$ under   the  symmetric  group $\mathcal{S}_N$ play the key role  \cite{Ou2006,SU3,NDBS,Tichy1,PartIndist,Rohde}, e.g., particles of  one species with    $\rho^{(int)}$  antisymmetric  under permutations emulate  behavior of the other species  \cite{FBunch,BSwithF}.  The probability formula    of an output configuration $\mathbf{m}=(m_1,\ldots,m_M)$  \cite{NDBS,PartIndist}   (applicable also   to fermions, see the Appendix) reads 

\be
\hat{p}(\mathbf{m})  = \frac{1}{\prod_{l=1}^M m_l!} \sum_{\tau,\sigma\in\S_N}J(\tau\sigma^{-1})
  \prod_{\alpha=1}^N U^*_{k_{\tau(\alpha)},l_\alpha}U_{k_{\sigma(\alpha)},l_\alpha},
\en{pm}
where     $l_1,\ldots,l_N$ are   output modes with multiplicities $(m_1,\ldots,m_M)$, and  a  complex-valued function $J(\sigma)$ of $\sigma\in \S_N$ is  defined as   
\be
J(\sigma) =    \vare(\sigma) \mathrm{Tr}\bigl( \rho^{(int)}  P_{\sigma}\bigr),\; \vare(\sigma) = \left\{\begin{array}{cc}1, & \mathrm{Bosons},\\
\mathrm{sgn}(\sigma),&\mathrm{Fermions}, \end{array} \right.
\en{J}
where $ P_\sigma\!\prod_{\alpha=1}^N{\!}^{\otimes}|j_\alpha\rangle = \prod_{\alpha=1}^N{\!}^{\otimes}|j_{\sigma^{-1}(\alpha)}\rangle$
is an   operator representation    of  $\sigma$ in the Hilbert space $\mathcal{H}^{\otimes N}$. 
  
 Identical particles  are called completely indistinguishable if  $\rho^{(int)}$  is  symmetric under   permutations,  $J^{(id)}(\sigma) = \vare(\sigma)$ (e.g., particles in  the same internal state), whereas particles with  orthogonal internal states are distinguishable,   $J^{(d)}(\sigma) = \delta_{\sigma,I}$   (see also Refs. \cite{PartIndist,BSwithF}).    The trace of   $\rho^{(int)}$    in   the symmetric subspace  $S_N\mathcal{H}^{\otimes N}$ with  $S_N=  (1/N!)\sum_\sigma  P_\sigma$,    i.e., $d(J) \equiv   \mathrm{Tr}\{S_N \rho^{(int)}\} = \frac{1}{N!}\sum_{\sigma\in\S_N}\vare(\sigma)J(\sigma)$,
is a suitable  measure of the  partial indistinguishability   \cite{Tichy1,TB} and   also  in other quantum information  applications  of  identical particles   \cite{Dist}.

Note that  $J(\sigma)$ of Eq.~(\ref{J}) is a p.s.d. function of $\sigma\in \mathcal{S}_N$, i.e.,   for any complex-valued  function  $z(\sigma)$  we have $\sum\limits_{\sigma_1,\sigma_2} z^*(\sigma_1) J(\sigma_1\sigma_2^{-1})z(\sigma_2) \ge 0
$, while $J(I)=1$ for  the  identity permutation $I$.  Therefore, there is  a   factorizing function  $\theta(\sigma)$ such that (see the Appendix)
\be
J(\sigma) = \sum_{\tau\in\S_N} \theta^*(\tau)\theta(\tau\sigma), \quad \sum_{\sigma\in\S_N} |\theta(\sigma)|^2 = 1.
\en{factJ}
Importantly,  \textit{any} normalized p.s.d.  function $J(\sigma)$  corresponds  to an input state of single  particles (see the Appendix), i.e., the whole   set of input states with  $N$ single particles  can be equivalently represented  by the  whole  convex set of  normalized p.s.d. functions $\mathcal{S}_N\to \mathbb{C}$.     

\textit{Probability to detect  all  particles in  $\K$ output modes.--}
The probability  for all  $N$ input particles  to gather  in  $\K$ output modes,  say  ${1,\ldots,\K}$ as in Fig. \ref{F0}(b), reads $p_N(J) = \sum_\mathbf{m}^\prime\hat{p}(\mathbf{m})$ where $\mathbf{m} = (m_1,\ldots,m_\K,0\ldots,0)$.  Defining  an $N$-dimensional p.s.d. Hermitian matrix  $H$, built from  the  submatrix  of $U$  on the rows $k_1,\ldots,k_N$ and columns $1,\ldots,\K$, and  the corresponding  $(N!)$-dimensional Schur power matrix   $\Pi(H)$ indexed by elements of $\mathcal{S}_N$,
 \be
 H_{\alpha,\beta} \equiv \sum_{l=1}^{\K} U_{k_\alpha,l}U^*_{k_\beta,l},\quad  \Pi_{\sigma,\tau}(H) =  \prod_{\alpha=1}^N H_{\sigma(\alpha),\tau(\alpha)},
 \en{H}
 we obtain  $p_N (J)$   in the form (see the Appendix) 
\be
 p_N(J)  = \sum_{\sigma\in\S_N} J(\sigma)   \Pi_{I,\sigma} 
= \sum_{\tau, \tau^\prime\in\S_N} \theta^*(\tau) \Pi_{\tau,\tau^\prime}\theta(\tau^\prime). 
\en{pN}
Therefore, $p_N(J)$   is  a convex combination of   eigenvalues  of $\Pi(H)$.   For the completely indistinguishable particles    $\theta^{(id)}(\sigma) =  {\vare(\sigma)}/{\sqrt{N!}}$, whereas for distinguishable  particles    $\theta^{(d)}(\sigma) = \delta_{\sigma,I}$.  Eq. (\ref{pN}) gives for these ideal cases:
\be
p^{(B)}_N    = \mathrm{per}(H),\; \; p^{(F)}_N  =  \mathrm{det}(H),\; \; p^{(d)}_N = \prod_{\alpha=1}^N H_{\alpha,\alpha},
\en{BF}
for bosons, fermions, and classical particles, respectively. Well-known inequalities  (see the Appendix) result in the following order 
\be
p^{(F)}_N\le p^{(d)}_N \le p^{(B)}_N.  
\en{Order}
Eq. (\ref{Order}) seem to suggest that $p^{(B)}_N$ and $p^{(F)}_N$ are the absolute maximum and minimum of $p_N(J)$. To establish such a property one has to show that they are the \textit{unique maximum (minimum) eigenvalues} of $\Pi(H)$.  A famous result of Schur \cite{Schur} states  that  the smallest eigenvalue of $\Pi(H)$ is   $\mathrm{det}(H)$, hence, the generalized antibunching is an universal attribute of the completely indistinguishable fermions (a unique minimum  for $\K\ge  N$). On the other hand,   the maximum eigenvalue of $\Pi(H)$ is generally unknown.  The permanent-on-top conjecture  (POT),  stating that  universally  it is  $\mathrm{per}(H)$,         proven   for  $N\le 3$ \cite{POT3},  has turned out  false   for $N\ge 5$ \cite{POTfalse}.  
  
Thus,  conditions on input state and/or network are needed to  ensure the maximum  probability being attained only by the completely indistinguishable bosons.   For  $\K=1$ Eq.~(\ref{pN}) gives    \mbox{$p_N(J) =  d(J) N!  \prod_{\alpha=1}^N|U_{k_\alpha,l}|^2$} with   the maximum for an arbitrary network  at $d(J) = 1$ or  $J(\sigma) = 1$ (see also  Refs. \cite{GenBunch,Tichy1,TB}).  Numerical  simulations~\footnote{Numerical simulations are  limited  to $N\le 7$ due to a large dimension of the Schur power matrix.} with  random p.s.d. Hermitian  matrices    reveal that  $p^{(B)}_N= \mathrm{per}(H)$ is not the maximum probability  \textit{only} for  $N\ge 5$ particles in    $2\le \K\le N-1$  output modes   (such a  state of $N\ge 5$ bosons is  necessarily a state of   non-classically correlated    particles, see below). 
More importantly, when $\K\ge N$,  a \textit{unique eigenvector}  $\theta^{(id)}_B(\sigma) = 1/\sqrt{N!}$ corresponds to $\mathrm{per}(H)$, i.e.,  the  maximum probability of $N$ particles to be detected  in   $\K\ge N$ output modes is attained  \textit{only} by the  completely indistinguishable bosons. Thus, bosons show  the generalized bunching property in $\K\ge N$ output modes.

For  only classically correlated bosons,  a unique maximum of $p_N(J)$  is attained only by the completely indistinguishable ones for all  $1\le \K\le M-1$.   Indeed, a classically-correlated    internal state can be expressed as  a convex combination of pure states, i.e.,    $\rho^{(int)} = \sum_{\mathbf{j}} \nu_{\mathbf{j}} |\phi_{j_1}\rangle \langle\phi_{j_1}|\otimes \ldots \otimes   |\phi_{j_N}\rangle \langle\phi_{j_N}|$ for  some arbitrary states $|\phi_{j_\alpha}\rangle\in\mathcal{H}$ and $ \nu_{\mathbf{j}} >0$,  $\sum_\mathbf{j}\nu_\mathbf{j} = 1$.  The corresponding  $J$-function reads
\be
 {J^{(cc)}}(\sigma) = \sum_{\mathbf{j}} \nu_{\mathbf{j}} \prod_{\alpha=1}^N \langle \phi_{j_{\sigma(\alpha)}}|\phi_{j_\alpha}\rangle. 
 \en{Jcc}
Setting $G^{(\mathbf{j})}_{\alpha,\beta} \equiv \langle \phi_{j_\beta}|\phi_{j_\alpha}\rangle$, we get  from Eqs. (\ref{H}), (\ref{pN}),  and  (\ref{Jcc})  
\be
p_N({J^{(cc)}}) =  \sum_{\mathbf{j}} \nu_{\mathbf{j}}\, \mathrm{per}(H\cdot G^{(\mathbf{j})}),
\en{pJcc}
 where  the dot stands for the Hadamard (by-element) product.   Thus, the permanental version of  Oppenheim's inequality  \cite{BSconj}, stating that for    two p.s.d. Hermitian matrices $H$ and $G$ (for  $G_{\alpha,\alpha}=1$), $
\mathrm{per}(H\cdot G) \le \mathrm{per}(H)$ would imply  the claimed  result in this case. Using    matrices $H$ that violate  the POT conjecture, it was checked that $\mathrm{per}(H\cdot G) < \mathrm{per}(H)$ for   any  $N$ random  states $|\phi_1\rangle,\ldots,|\phi_N\rangle$, with  at least two   linearly independent.  
 
\textit{Assessment  protocol of a BS  device.--}  The average $\langle p_N(J)\rangle$ over  Haar-random   networks gives an idea of quantitative features of the generalized bunching/antibunching effect.  By the unitary invariance of the Haar measure,  the average probability $\langle p_N^{(B,F)}\rangle$    depends only on the ratio of  considered output configurations (see the Appendix): 
\be
  \langle{p}^{(B,F)}_N\rangle  = \frac{\K(\K\pm 1) \cdot\ldots\cdot(\K \pm N \mp1)}{M(M\pm1)\cdot\ldots\cdot(M\pm N \mp1)},
 \en{pNHaar}
here (and below)  the upper  (lower) signs  stand  for bosons (fermions).  For distinguishable particles there is no exact result, but  for  $M\gg 1$  it can be shown that  (see the Appendix)  
\be
 \langle{p}^{(d)}_N\rangle = \left( \frac{\K}{M}\right)^N\left[1+O\left(\frac{N^2}{\K M}\right)\right]. 
\en{classAs}
The  average probability ratio becomes 
\be
\frac{ \langle{p}^{(B,F)}_N\rangle }{ \langle{p}^{(d)}_N\rangle }=\left[1+O\left(\frac{N^2}{\K M}\right)\right]\prod_{i=1}^{N-1}\frac{1\pm l/\K}{1\pm l/M}.
\en{Brat}

For   $NL\ll M$, where $L=M-\K$,   the   detection probability is close to $1$: $\langle{p}^{(B,F)}_N\rangle=\prod_{l=0}^{N-1}[1-L/(M\pm l)]= 1- O(LN/M)$, whereas   the r.h.s. of Eq.~(\ref{Brat}) gives $ \langle{p}^{(B,F)}_N\rangle /\langle{p}^{(d)}_N\rangle  \approx 1 \pm  {LN(N-1)}/(2M^2)$.   In this case one needs   $R\gg   M^4/(N^4L^2)$ runs for   the  ratio  (\ref{Brat}) to surpass  the  statistical error $O(1/\sqrt{R})$ in  experimental data. But the ratio in  Eq. (\ref{Brat}) is attained \textit{only} by the completely indistinguishable bosons (fermions), hence,  it is \textit{a reliable   witness}, detectable  in polynomial number of runs,   of their  complete indistinguishability during propagation, i.e., that  no  decoherence process  has  contributed to   output statistics.     
Therefore,   we have   an efficient    protocol  for assessment   of a BS device with an \textit{arbitrary} network, Fig.~\ref{F0}.  The only known protocol   for a BS device with an arbitrary  network  \cite{AA2}, experimentally verified \cite{E6}, discriminates  the BS  and  uniform   distributions.  Our protocol  discriminates against \textit{all}  other than BS  samplers, which are physically realizable with particles in a linear network,  including the classical ($\mathcal{M}_A$), the fermion ($\mathcal{F}_A$),  the random-classical ($\mathcal{B}_A$) samplers of Ref.   \cite{AA2},  and     the  random-phase bosons \cite{ZlawBS,StatB}.

The assessment  protocol has two stages. At stage (I),  Fig.~\ref{F0}(I),  by using photon-number resolving detectors (e.g.,  by cascading bucket detectors \cite{PRD}) one checks  that  sources  produce   $N$  single photons.  At stage (II),  Fig.~\ref{F0}(II),  employing  the universality of   generalized bunching,  one verifies    experimental statistics against  the probability $p^{(B)}_N$ (\ref{BF}) using $L= M-\K$ bucket detectors.

The protocol  requires \textit{only one}  matrix permanent ${p}^{(B)}_N=\mathrm{per}(H)$ of $H$ in Eq. (\ref{H}) (a single set of $\K$ modes is used)  to an  error $\epsilon=O(N^{-\kappa})$ for some $\kappa>0$ (statistical error in experimental data for a polynomial number of runs). For  $N\gg 1$ in the dilute limit  $M = O(N^{2+\delta})$ with  $\delta>0$ and $L = O(N)$  it can be shown (using that we select $\K$ modes arbitrarily) that  \textit{only polynomial  in $N$ computations} $\mathcal{C}_N$   are required  (see the Appendix; in this case $ \langle{p}^{(B)}_N\rangle = 1 - O(N^{-\delta})$ and  $ \langle{p}^{(B)}_N-{p}^{(d)}_N\rangle   = O(N^{-1-2\delta})$).  One can estimate that, on average  over all choices of $\K$ output  modes, $\mathcal{C}_N= O(N^{\frac{\kappa+1}{\delta}})$ (e.g., setting $\kappa = 2[1+\delta]$ allows one to distinguish  the quantum and classical cases). 

The protocol applies also to  the Scattershot BS \cite{SCBS}, recently experimentally tested \cite{E7},  which uses  heralded single photons in $N$ random  input modes  in each run. The probability  describing experimental statistics of a Scattershot BS is  well approximated by   $\langle p^{(B)}_N\rangle$ (\ref{pNHaar}) already for a few hundred runs of such a  device (see Fig. \ref{F1} below), i.e., \textit{no computations required.}

Stage (I) is designed to  expose  all attempts  to bypass the  universality property using  inputs with variable number of particles per mode (not required under certified input).  Indeed, an input  with   any  distribution of the completely indistinguishable  bosons between $M$ input modes
\be
\rho = \sum_{\mathbf{n}} p_\mathbf{n}|\mathbf{n}\rangle\langle\mathbf{n}|, \quad p_\mathbf{n}\ge 0,\quad \sum_\mathbf{n} p_\mathbf{n} =1,
\en{contest}
where   $n_1+\ldots+n_M=N$, has the  Haar-average probability equal to  $\langle p^{(B)}_N\rangle$ (see the Appendix).  For example, take   the  random-phase bosons of Refs. \cite{ZlawBS,StatB}, i.e., an input    where  each boson is  in a coherent superposition  of $S$ input modes (and in an internal state $|\phi\rangle$) described by operator $A (\mathbf{\theta}) =S^{-\frac12}\sum_{j=1}^S  e^{i\theta_j}a_{k_j,\phi}$ with random phases $\theta_1,\ldots,\theta_S$     \cite{ZlawBS,StatB}. The  density matrix  of this input reads
\begin{eqnarray}
\label{rphs}
 && \rho_{s} =  \frac{(S-1)! S^N}{(S+N-1)!} \prod_{j=1}^S\int\limits_0^{2\pi}\frac{\rd \theta_j}{2\pi}   \left[A^\dag(\mathbf{\theta})\right]^N |0\rangle \langle0|\left[A(\mathbf{\theta})\right]^N\nonumber\\
 && = \frac{N!(S-1)!}{(S+N-1)!}\sum_{\mathbf{n}} |\mathbf{n}\rangle\langle\mathbf{n}|,
\end{eqnarray}
where  $\mathbf{n} = (n_{k_1},\ldots,n_{k_S})$, $n_{k_1}+\ldots+n_{k_S}=N$.  A  source of   $\rho_s$  with  $S=N$ is exposed  at stage (I)  by  a  vanishing   probability of an input with one particle per occupied mode,      for  $N\gg 1$  scaling as $\sim 4^{-N}$.  Stage (I) exposes also the sampler   $\mathcal{B}_A$ of Ref. \cite{AA2}, a ``mockup distribution of BS"  physically realized by  distributing $N$ uncorrelated  distinguishable particles   randomly over  $N$   input modes,  with the  probability of such a particle to land in an  output mode $l$ being $p(l) = \frac{1}{N}\sum_{\alpha=1}^N |U_{k_\alpha,l}|^2$.  The probability of single particles at input is   $N!/N^{N}\sim e^{-N}$. 

\squeezetable\begin{table}[h!]
\centering
\begin{tabular}{ |c| c c c  c c c c  c c c c  c c c c  c c c | } 
 \hline 
$N$ & 3  &   4 &    5 &    6  &   7  &   8 &    9   & 10  &  11 &   12 &   13 &   14  &  15   & 16   & 17   & 18  &  19 &   20 \\    \hline 
 $L$ &   2 &     2&     3 &    4 &    5 &    5  &   6 &    7  &   7  &   8  &   9    & 9  &  10  &  11  &  11  &  12  &  13  &  14 \\    \hline 
 $M$ &    5 &     8&    13&    18&    25&    32&    41 &   50 &    61 &   72 &    85 &   98 &  113 &  128 &  145 &  162 &  181&   200\\
   \hline
\end{tabular}
 \caption{Network size $M$  and   $L = M-\K$ as functions of  $N$.  Here   $\K$  is selected by maximizing the ratio of  Eq. (\ref{Brat})  under the condition that $\langle p^{(B)}_N\rangle \ge 0.25$ (note that $\K\ge N$). }
\label{table1}
\end{table}
\begin{figure}[htb]
\begin{center}
\includegraphics[width=0.45\textwidth,height=0.325\textwidth]{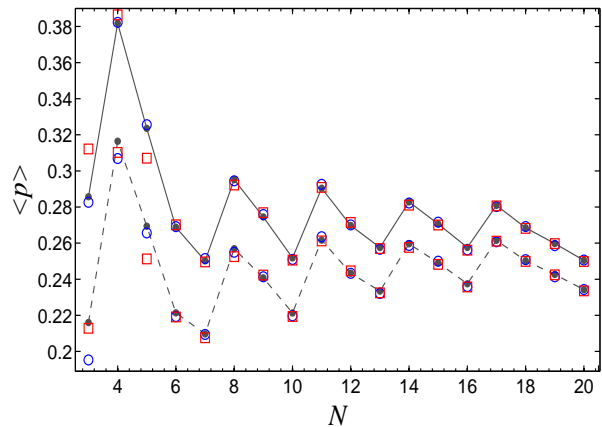}
\caption{(Color online)  The analytical  quantum average probability  $\langle p^{(B)}_N\rangle$ Eq. (\ref{pNHaar})    (dots  on the solid line)  and   the  approximation Eq. (\ref{classAs}) of the classical one  $\langle p^{(d)}_N\rangle$    (dots on the dashed line) vs.   numerical averaging  with    $1000$ 
Haar-random networks (circles).  The   squares give  the  quantum and classical probability  for the Scattershot BS  (estimated with $500$ runs)    with a randomly chosen  network for each value of $N$.  }\label{F1}
\end{center}
\end{figure}
Fig. \ref{F1}  gives numerical results, where  $M = [\frac12N^2]$ (integer part), and $L= M- \K$  is obtained by   maximizing the ratio (\ref{Brat}) for $\langle p^{(B)}_{N}\rangle\ge 0.25$ (see Table \ref{table1}).
 
 The distinguishability error $1-d(J) \approx   (1-F)(N-1)$ \cite{TB}  in a BS device  ($F$   is the mean fidelity of  indistinguishability of photons)  can be  assessed    from  an experiment using the first-order approximations (see the Appendix):
\begin{eqnarray}
\label{Slaw}
&& \!\!\!\! p^{(B)}_N - p_N(J) \approx  \frac{1-d(J)}{N-1}\left[Np^{(B)}_N -  \frac{d}{dx}\mathrm{per}\{H(x)\}_{x=1} \right],\nonumber\\
&& \!\!\!\! \langle  p^{(B)}_N\rangle  - \langle p_N(J) \rangle \approx    [1- d(J)]\frac{N}{M} \langle p^{(B)}_{N-1}\rangle,
\end{eqnarray} 
where $H_{\alpha,\beta}(x) \equiv \delta_{\alpha,\beta}H_{\alpha,\beta} x + (1-\delta_{\alpha,\beta})H_{\alpha,\beta}$ ($\mathrm{per}\{H(x)\}$ is a  polynomial  in  $x$ of order $N$). The second law  is valid for  $M\gg N^2$ and applies  to  the Scattershot BS. 
 
Difference between an experimental and the theoretical  probability $p^{(B)}_N - p^{(exp)}_N$  would reflect presence of other errors in a device.  How the BS regime is affected by errors in a network matrix  can be estimated beforehand.  The aim of such a certification, besides eliminating the possibility of  loopholes (see the Appendix),  is to guarantee that network  errors would be below an acceptable  level in the  BS regime. The theoretical basis is provided in Refs. \cite{LP,Ark}. Besides an unwanted distinguishability and network matrix errors, errors reported in BS experiments \cite{E1,E2,E3,E4,E5,E6,E7} include  higher-order photon numbers,  estimated   at  stage (I) of our protocol, and the photon losses, which can be    directly estimated   at a network output (see also  Ref. \cite{BSscal}).   A BS  device with a   fixed-ratio  of  lost  photons is believed to be hard to simulate  on a classical computer  \cite{AA,RR} with some progress in proof  \cite{AB}.  Since a  lossy linear $M$-mode network is equivalent to an $2M$-mode unitary  one, with one half of  output modes being  unaccessible ``loss channels"  (see the Appendix),   our assessment  protocol  applies also to linear lossy networks. In this case in  Eq. (\ref{H}) the proper non-unitary network matrix  $U$ must be used. It can be experimentally characterized  with only   classical coherence \cite{White}.  
 
\textit{Conclusion.--}    We have discovered the generalized boson bunching and fermion antibunching in linear networks and proposed an assessment protocol for BS verifying in a \textit{polynomial} number of experimental runs that  a BS  device  with a random  linear  network  operates at the full $N$th-order quantum coherence compatible only with $N$ completely indistinguishable bosons, i.e., the very  physical origin of its   quantum supremacy.  The protocol requires  \textit{only polynomial} classical computations for the standard version of BS, whereas for     the Scattershot  version  (with better prospects for scalability)  \textit{analytical} results are available.  In general terms, the generalized bunching is a generalization of  the famous HOM effect,  revealing the complete indistinguishability  of $N$ photons in an arbitrary (nontrivial) $M$-mode  network, which may find other   applications whenever linear quantum networks are used.

\textit{Acknowledgement.--} This work was supported by the CNPq (Brazil) under grants 304986/2011-9 and 304129/2015-1.

\newpage

 \begin{widetext}
\begin{center}
{\large \textbf{Appendix}}
\end{center}
\end{widetext}

\section{An efficient assessment protocol for BS on an uncertified network matrix  allows for loopholes}

Verification of Boson Sampling  (BS) with a random network  without knowledge of a network matrix is an ill-defined problem \cite{Gogolap,AA2ap}. 
In view of this, a natural  question arises: Is it somehow possible to verify the full $N$th order quantum coherence  compatible only with  $N$  completely indistinguishable  bosons and a (special) network matrix simultaneously \textit{in a polynomial number of runs} in  the same BS regime on such a device?  This may seem indeed possible due to  presence of a  high symmetry in a network matrix, leading to very distinct features in output distribution. Let us call such an assessment of BS  \textit{the holistic assessment}. 

However, careful analysis reveals loopholes in an  efficient assessment protocol if the latter is used as a holistic one.    Let us analyze in detail   an  assessment  protocol for BS based on the  Fourier network    \cite{ZlawBSap} (a highly symmetric Bell multiport), where a large fraction  of the output configurations are forbidden by  symmetry  for completely indistinguishable bosons   \cite{ZlawBSap,MPBFap}.  Let us recall the necessary  details  of the assessment protocol  proposed in Ref. \cite{ZlawBSap}. One considers Fourier network with $M = N^p$  modes (where $p \ge 2$), i.e., given by the matrix
\be
F^{(M)}_{k,l} = \frac{1}{\sqrt{M}}\exp\left( i\frac{2\pi}{M}kl\right), \quad k,l = 1,\ldots, M. 
\en{F1} 
For single particles in the input  modes $1\le k_1,\ldots,k_N\le M$ with the following  cyclic symmetry 
\be
k_{\alpha+1} = k_\alpha + N^{p-1}
\en{cyclic} 
where, obviously, $1\le k_1 \le N^{p-1}$, only those output configurations $1\le l_1,\ldots,l_N\le M$ which satisfy
\be
\sum_{\alpha=1}^N l_\alpha = q N, \quad q \in \mathcal{N}, 
\en{F2}
i.e.,  have  the sum of mode indices divisible by $N$, are realized with completely indistinguishable bosons in such a network \cite{ZlawBSap}. Moreover, the number of realized output configurations is  only a  fraction $1/N$ of the total number of all  \textit{a priori} possible ones, which allows efficient verification of the full $N$th order indistinguishability by counting the number of violations of the forbidden events, on the order of  $1- 1/N$  of the total number of events \cite{ZlawBSap}. 

The above described protocol is subject to loopholes, if it is understood as a holistic one, i.e. when  no certification of a network matrix  \textit{in a different regime} is performed~\footnote{The  authors of Ref. \cite{ZlawBSap}  do not  claim that their assessment  protocol is understood as a holistic one, it is used only to illustrate appearance  of serious loopholes  if one considers  the protocol as holistic.}. Loopholes appear due to a combined effect of network and distinguishability errors.  Let us illustrate this by exposing a loophole in  the following case: $N = 2N_1$ and $p=2$ (for simplicity). Denote $M_1 = N_1^2$. Since no information on a network is assumed, instead of  $F^{(M)}$ (\ref{F1}) one cannot rule out   the following one ($P$ is a permutation)
\be
U =  P\left[F^{(M_1)} \oplus F^{(M_1)} \right]P^\dag\oplus  P \left[ F^{(M_1)}  \oplus F^{(M_1)} \right]P^\dag, 
\en{F3}
where $U$ is  a block-structured matrix with two diagonal  blocks, each containing two  $M_1$-dimensional Fourier matrices (\ref{F1}) whose rows and columns are permuted by  $P$      in such a way that  each odd  (even) row and column in the two blocks  of $U$  of size $2M_1$ corresponds to one  and the same submatrix $F^{(M_1)}$.  

Consider now $N$ bosons  divided into  two groups of $N_1$  completely indistinguishable bosons, whereas  bosons from different groups being  distinguishable (e.g., bosons in group $i=1,2$ are in an internal state $|\phi_i\rangle$ and $\langle \phi_1|\phi_2\rangle=0$). Assume that such bosons are launched into $U$ (\ref{F3}) observing the cyclic symmetry (\ref{cyclic}), where  the bosons in the first  $N_1$ input  modes of $U$ are  from the same group (i.e., completely indistinguishable).  Since $N$ is even, the parity of an input mode  $k_\alpha$ of $U$  is the same as that of $ k_1$. Thus, the   first  (second) group of $N_1$  completely indistinguishable bosons are launched in one of the two  submatrices $F^{(M_1)}$ of  the respective block,  in a similar cyclic symmetry  as in  Eq. (\ref{cyclic}) for its proper mode indices and $N$ substituted by $N_1$.  By Refs.~\cite{MPBFap,ZlawBSap}   and the block-structure of $U$, the allowed  output configurations  must  satisfy  conditions similar to Eq. (\ref{F2}) now for   the proper output indices, say, $1\le l^{(i)}_1,\ldots,l^{(i)}_{N_1}\le M_1$, of that  particular submatrix $F^{(M_1)}$, i.e., 
\be
\sum_{\alpha=1}^{N_1} l^{(i)}_\alpha = q^{(i)}N_1, \quad q^{(i)} \in \mathcal{N},\quad i=1,2.
\en{F4}

Let us now  verify that  for a cyclic input (\ref{cyclic}) \textit{all} allowed output configurations in a network $U$ (\ref{F3}), with the above described partially distinguishable bosons,  belong to  the allowed set $\mathcal{A}$ of  the Fourier network  $F^{(M)}$ (\ref{F1}) with completely indistinguishable bosons  (where there is an exponential number of such  $|\mathcal{A}|\sim N^{N-1}$).  First, let us assume  that $k_1$ is even.  The   indices $1\le l_1,\ldots,l_N\le M$ of output modes of  $U$, where  bosons from one of the two groups    can end up, are derived from the output indices in the corresponding submatrix $F^{(M_1)}$, satisfying Eq. (\ref{F4}), by the following two rules (respectively, for the first and second blocks  of $U$)
\be
l_\alpha = 2l^{(1)}_\alpha,  \; l_{N_1+\alpha} = 2l^{(2)}_\alpha + 2M_1, \; \alpha =1,\ldots,N_1. 
\en{F5}
Since $N = 2N_1$   Eq. (\ref{F2}) is satisfied by such output mode indices
\be
\sum_{\alpha=1}^N l_\alpha  = (q^{(1)} + q^{(2)} + M_1)N. 
\en{F6}
Similarly, when $k_1$ is odd we have a similar relation for the output mode indices
\be
l_\alpha = 2l^{(1)}_\alpha-1,  \; l_{N_1+\alpha} = 2l^{(2)}_\alpha - 1 + 2M_1, \; \alpha =1,\ldots,N_1,
\en{F5}
with a similar  conclusion.  Thus, all of the exponentially many  allowed output configurations  in $U$ belong to $\mathcal{A}$. Though we have considered an even number of bosons, a similar example can be devised for an odd number as well.  

The considered combination of a   matrix $U$ and partially distinguishable bosons is just a  special simple  case of many such  combinations of  a network and partially distinguishable bosons (see  also   Ref. \cite{PartIndistap}) with the allowed output configurations  being  an exponential subset of  $\mathcal{A}$. For example,  network $U$ (\ref{F3})  relates its  output modes to input modes of the same parity,  but  a slightly modified  network consisting of $U$ (\ref{F3}) followed by a ``noisy" network implementing a random  shift of  all   modes  by  $1$  (mod $M$) with probability $1/2$  will have an exponential number of allowed output configurations all belonging to $\mathcal{A}$  and no  parity symmetry~\footnote{To  rule out such and other possible  networks is precisely  the point in an  \textit{independent} certification of a network.}.  Due to many such possibilities, there is  no way to guess, not knowing a network matrix  $U\ne F^{(M)}$,  what  kind of  particular feature distinguishes  an exponential in $N$ subset of $\mathcal{A}$  which corresponds to such $U$  and partially distinguishable bosons. An experimentalist  assuming   that a network matrix is   close to $F^{(M)}$ (\ref{F1}) and  input  bosons close  to being completely indistinguishable,  attempting  a holistic assessment by the above protocol, runs   only  the  BS regime.  After a polynomial number of runs of a device, which in reality has  one  of the  alternative  possibilities of network matrix  $U\ne F^{(M)}$ and partially distinguishable bosons,    the  experimentalist would conclude that the input is \textit{fully}  compatible with the completely indistinguishable bosons  and the matrix with  $F^{(M)}$, since with only  a polynomial number of runs one cannot tell from an experimental statistics  that only some exponential in $N$ part of $\mathcal{A}$ is actually realized.

The conclusion is that, to close all possible  loopholes in an efficient assessment protocol of BS,   it is  absolutely necessary to \textit{independently} certify a network matrix, e.g., using an efficient method  of Ref. \cite{Whiteap} requiring   only classical coherence. Besides eliminating the possibility of  loopholes,  such a certification would also  guarantee that network  errors are below  an established acceptable  level  when   BS is run  on such a network. The theoretical basis for a network  assessment is provided in Refs. \cite{LPap,Arkap}, where  the  effect of  network matrix  errors  on  output probability distribution of a BS device with such a network  is estimated.

\section{Probability of $N$  particles to gather  in $\K$ output modes of a linear network} 
 \label{sec2}

Let us first recall principal  steps in derivation of the  output  probability  distribution of  $N$ identical bosons/fermions at input  of a linear network $U$ (for bosons  this result  was derived in   Refs. \cite{NDBSap,PartIndistap}).   We will consider simultaneously both species (in case of fermions an order  of  creation and annihilation  operators  is assumed).   A more general input   is assumed, with $0\le n_k\le N$ particles per  input mode $k$ (fermions have    linearly independent internal states in each input mode). A general input state of configuration $\mathbf{n} = (n_1,\ldots,n_M)$  reads 
 \be
 \rho(\mathbf{n}) = \sum_{i}q_i|\Psi_i\rangle\langle\Psi_i|,
 \en{A01}  
 with  $q_i\ge 0$, $\sum_i q_i =1$,  and 
\be
|\Psi_i \rangle = \frac{1}{\sqrt{\mu(\mathbf{n})}}\sum_{\mathbf{j}}C^{(i)}_{\mathbf{j}}\prod_{\alpha=1}^N a^\dag_{k_\alpha,j_\alpha}|0\rangle,
\en{A02}
where  a basis state $|j\rangle \in\mathcal{H}$ in the  internal space is introduced  (e.g., a basis  function  of  spectral shape of a photon),  $\mathbf{j} = (j_1,\ldots,j_N)$ and $\mu(\mathbf{n}) = \prod_{k=1}^Mn_k!$. Permutation symmetry (anti-symmetry) of  creation operators for bosons (fermions)  allows to chose  expansion coefficients $C^{(i)}_{\mathbf{j}}$    symmetric     (anti-symmetric) with respect to the Young  subgroup $\mathcal{S}_{\mathbf{n}}\equiv \mathcal{S}_{n_1}\otimes \ldots \otimes \mathcal{S}_{n_M}$ of the symmetric group $\mathcal{S}_N$, where $\mathcal{S}_{n_k}$ corresponds to permutations of  internal states of particles    in   input mode $k$ between themselves.   Such  coefficients are normalized by  $\sum_{\mathbf{j}} |C^{(i)}_{\mathbf{j}}|^2 =1$.

The probability   of an output configuration $\mathbf{m} = (m_1,\ldots,m_M)$  is given as  \cite{NDBSap,PartIndistap,BSwithFap}
\be
\hat{p}(\mathbf{m}|\mathbf{n}) = \mathrm{Tr}( \rho(\mathbf{n}) \mathcal{D}(\mathbf{m})),
\en{A1}
where $\rho$ is the input state Eq. (\ref{A01}) and $\mathcal{D}(\mathbf{m})$ is the  detection  operator \cite{PartIndistap,BSwithFap} ($|0\rangle$ is  Fock vacuum state)
\be
 \mathcal{D}(\mathbf{m}) =   \frac{1}{\mu(\mathbf{m})} \sum_{\mathbf{j}} \left[\prod_{\alpha=1}^N b^\dag_{l_\alpha,j_\alpha}\right]|0\rangle\langle0|\left[\prod_{\alpha=1}^N b_{l_\alpha,j_\alpha}\right]. 
\en{A2}
One can evaluate the trace in Eq. (\ref{A1})  by first expressing the input mode operators in Eq. (\ref{A01}) through the output ones using $a^\dag_{k,j} = \sum_{l=1}^M U_{k,l} b^\dag_{l,j}$ and then employ the  following  identity (see also Refs. \cite{NDBSap,PartIndistap,BSwithFap})
\begin{eqnarray}
\label{A3}
&&\langle0|\left[\prod_{\alpha=1}^N b_{l_\alpha,j_\alpha}\right]\left[\prod_{\alpha=1}^N b^\dag_{l^\prime_\alpha,j^\prime_\alpha}\right]|0\rangle 
\nonumber\\
&&= \sum_{\sigma\in\S_N}\vare(\sigma)\prod_{\alpha=1}^N  \delta_{l^\prime_\alpha,l_{\sigma(\alpha)}} \delta_{j^\prime_\alpha,j_{\sigma(\alpha)}}. 
\end{eqnarray}
Substituting Eq. (\ref{A01}) and (\ref{A2})  into Eq. (\ref{A1}) and using Eq. (\ref{A3}) in the two inner products one obtains   the probability of an output configuration $\mathbf{m}$ in a linear network $U$ in the form
\be
\hat{p}(\mathbf{m}|\mathbf{n})  = \frac{1}{\mu(\mathbf{m})\mu(\mathbf{n})} \sum_{\tau,\sigma\in \S_N}J(\tau\sigma^{-1})
  \prod_{\alpha=1}^N U^*_{k_{\tau(\alpha)},l_\alpha}U_{k_{\sigma(\alpha)},l_\alpha}.
\en{A4}
where $l_1,\ldots,l_N$ are   output modes, $1\le l_\alpha\le M$, with multiplicities $(m_1,\ldots,m_M)$, whereas   function  $J(\sigma)$   and the internal state are  
\be
 \rho^{(int)}  = \sum_i q_i |\psi_i\rangle\langle\psi_i|,\quad |\psi_i\rangle \equiv \sum_{\mathbf{j}} C^{(i)}_{\mathbf{j}}  \prod_{\alpha=1}^N{\!}^{\otimes}|j_\alpha\rangle
\en{Apint}
and 
\be
J(\sigma) =    \vare(\sigma) \mathrm{Tr}\bigl( \rho^{(int)}  P_{\sigma}\bigr),\; \vare(\sigma) = \left\{\begin{array}{cc}1, & \mathrm{Bosons},\\
\mathrm{sgn}(\sigma),&\mathrm{Fermions},\end{array} \right.
\en{AJ}
where $ P_\sigma\!\prod_{\alpha=1}^N{\!}^{\otimes}|j_\alpha\rangle = \prod_{\alpha=1}^N{\!}^{\otimes}|j_{\sigma^{-1}(\alpha)}\rangle$
is the   operator representation    of  $\sigma$ in $\mathcal{H}^{\otimes N}$. 
 Note that the permutation symmetry (anti-symmetry) of the input state (\ref{A01})-(\ref{A02})  for bosons (fermions), i.e., $P_\pi \rho^{(int)} = \rho^{(int)}P_\pi = \vare(\pi)\rho^{(int)}$ for any $\pi\in \mathcal{S}_{\mathbf{n}}$, implies that 
\be
J(\sigma\pi) = J(\pi\sigma) = J(\sigma), \quad  \forall \pi \in \mathcal{S}_{\mathbf{n}}. 
\en{A5}

A network input  consists of \textit{ completely indistinguishable}  bosons (fermions) if the corresponding $J$-function reads $J^{(id)}(\sigma) = \vare(\sigma)$ \cite{PartIndistap,BSwithFap}  (in case of fermions $\mu(\mathbf{n})=1$).  This case allows one to completely  neglect the internal degrees of freedom.  Probabilities at a network output are expressed through the usual matrix permanent and determinant, respectively.   The simplest case of   completely indistinguishable particles consists of    all particles being   in the same internal state $|\phi\rangle$, giving  $\rho^{(int)} = \left(|\phi\rangle\langle\phi|\right)^{\otimes N}$. 

The other limiting case, which may be identified as the \textit{classical case}, since the output probabilities are the same as in the case of classical particles, corresponds  to a    ``block-structured"  $J$-function  (see also Refs. \cite{NDBSap,PartIndistap})
\be
J^{(d)}(\sigma) = \sum_{\pi\in\mathcal{S}_{\mathbf{n}}} \delta_{\sigma,\pi}.
\en{A6}
Function $J(\sigma)$ of Eq. (\ref{A6})  appears   when the  internal states of identical particles from different input modes become  orthogonal: $\mathrm{Tr}\{\rho^{(int)}P_\sigma\} =0$ for $\sigma\notin  \mathcal{S}_{\mathbf{n}}$,  whereas (by the symmetry of $C_{\mathbf{j}}$) we always have  $\vare(\sigma)\mathrm{Tr}\{\rho^{(int)}P_\sigma\} =1$ for $\sigma\in  \mathcal{S}_{\mathbf{n}}$ (i.e., distinguishable particles from the same input mode cannot be discriminated by a linear network  from the completely indistinguishable bosons).  Note that the subgroup $\S_{\mathbf{n}}$ acts as identity on the indices $k_1,\ldots,k_N$ of matrix $U$ in Eq. (\ref{A4}), thus the sum over $\S_{\mathbf{n}}$ in Eq. (\ref{A6})  cancels $\mu(\mathbf{n})$ in the denominator in Eq. (\ref{A4}), resulting in the familiar formula for the probability in the classical case, expressed through the matrix permanent of  doubly stochastic matrix with elements $|U_{kl}|^2$. Obviously, for  single particles at input ($n_k\le 1$) we have $J^{(d)}(\sigma) = \delta_{\sigma,I}$. 

The total probability of detecting \text{all} $N$ input particles at a preselected (and fixed) set of $1\le \K\le M$ output modes, say, the first $\K$ modes,  is  a sum of  $\hat{p}(\mathbf{m}|\mathbf{n})$  (\ref{A4}) with $m_{\K+1} =\ldots m_M=0$. We have  
\begin{eqnarray}
\label{A7}
&& p_N(J) = {\sum_{\mathbf{m}}}^\prime\hat{p}(\mathbf{m}|\mathbf{n}) =      \frac{1}{N!} \sum_{l_1=1}^\K \ldots \sum_{l_N=1}^\K\nonumber\\
&&\times \frac{1}{\mu(\mathbf{n})}\sum_{\tau,\sigma\in\S_N}J(\tau\sigma^{-1})
  \prod_{\alpha=1}^N U^*_{k_{\tau(\alpha)},l_\alpha}U_{k_{\sigma(\alpha)},l_\alpha}\nonumber\\
&&=\frac{1}{\mu(\mathbf{n})}\sum_{\sigma^\prime\in\S_N} J(\sigma^\prime)  \prod_{\alpha=1}^N H_{\alpha,{\sigma^\prime(\alpha)}}= \frac{1}{\mu(\mathbf{n})} \sum_{\sigma\in\S_N} J(\sigma)   \Pi_{I,\sigma}\nonumber\\
&&  = \frac{1}{\mu(\mathbf{n})} \sum_{\tau, \tau^\prime\in\S_N} \theta^*(\tau)\theta(\tau^\prime) \Pi_{\tau,\tau^\prime}. 
 \end{eqnarray}
where we have transformed the  sum over output configurations $\mathbf{m}$    into that  over  output mode indices $l_1,\ldots,l_N$ with the combinatorial coefficient $ \mu(\mathbf{m})/N!$, defined the following matrices
\be
 H_{\alpha,\beta} \equiv \sum_{l=1}^{\K} U_{k_\alpha,l}U^*_{k_\beta,l},\;  \Pi_{\sigma,\tau}(H) =  \prod_{\alpha=1}^N H_{\sigma(\alpha),\tau(\alpha)},
 \en{AH}
 reordered the product $ \prod_{\alpha=1}^N H_{{\sigma(\alpha)},{\tau(\alpha)}}=  \prod_{\alpha=1}^N H_{\alpha,{\tau\sigma^{-1}(\alpha)}}$,  and  defined  $\sigma^\prime=\tau\sigma^{-1}$ (the sum over $\tau$ cancels  the factor $1/N!$). 
 
Note that for  a p.s.d. Hermitian $H$,  as a principal submatrix of the  tensor product matrix $H^{\otimes N}$ the Schur power matrix is a p.s.d. Hermitian as well.

Consider now the limit cases. In case of the completely indistinguishable particles      $\theta^{(id)}(\sigma) = \frac{\vare(\sigma)}{\sqrt{N!}}$ (since $J^{(id)}(\sigma) = \vare(\sigma)$).    Eq. (\ref{A7}) gives for bosons and fermions: 
\be
p^{(B)}_N =    \frac{ \mathrm{per}(H)}{\mu(\mathbf{n})}, \quad p^{(F)}_N =  \mathrm{det}(H)  \delta_{\mu(\mathbf{n}),1},
\en{A8}
where 
$
\mathrm{per}(A) \equiv \sum_\sigma \prod_{\alpha=1}^N A_{\alpha,\sigma(\alpha)}.
$
For  classical particles (e.g., distinguishable bosons or fermions from different input modes)   we get from Eq. (\ref{A6}) that 
\be
\theta^{(d)}(\sigma) = \frac{1}{\sqrt{\mu(\mathbf{n})}}\sum_{\pi\in\mathcal{S}_{\mathbf{n}}}\delta_{\sigma,\pi},
\en{A10}
which results in 
\be
p^{(d)}_N = \prod_{\alpha=1}^N H_{\alpha,\alpha}. 
\en{AClass}

The following order of the limit-case probabilities can be easily established
\be
p^{(F)}_N \le p^{(d)}_N\le  p^{(B)}_N, 
\en{A11}
valid for     \textit{arbitrary} number of particles  per input  mode.  
Indeed, for a p.s.d. Hermitian matrix $H$, which we rearrange in a block-matrix form
\be
H = \left(\begin{array}{cc} H^{(1,1)} & H^{(1,2)} \\
H^{(2,1)} & H^{(2,2)} \end{array}\right).
\en{A12}
the following inequality is known for the matrix determinant \cite{Fisher}:
\be
\mathrm{det}(H) \le \mathrm{det}(H^{(1,1)}) \mathrm{det}(H^{(2,2)}). 
\en{A13}
 Similarly,  for  the matrix permanent \cite{Lieb}
\be
\mathrm{per}(H) \ge \mathrm{per}(H^{(1,1)}) \mathrm{per}(H^{(2,2)}). 
\en{A14}
By repeated application of Eqs. (\ref{A13}) and (\ref{A14})  it is easy to demonstrate that Eq. (\ref{A11})   holds.

\section{Factorization of  $J$-function  and its representation   through  a density matrix  }

To show that the probability  $p^{(B)}_N$  ($p^{(F)}_N$) for the completely indistinguishable bosons (fermions)  corresponds to the absolute maximum (minimum) over \textit{arbitrary} input states of particles in a given configuration $\mathbf{n}$,  one has to know to   what class of functions the physical $J$-functions,  i.e., describing an  input  of a linear network, belong.  Let us show that any  normalized   by $J(I) = 1$  p.s.d. function $J(\sigma)$  can be represented in the form of Eq. (\ref{AJ})  with some state $\rho^{(int)}$ (\ref{Apint}).  Since   $\mathrm{sgn}(\sigma)J(\sigma)$ is also a normalized p.s.d function, it is sufficient to consider bosons, $\vare(\sigma) = 1$.  

Consider  a linear subspace $\mathcal{L}\subset \mathcal{H}^{\otimes N}$   defined as  the linear span of vectors $|\sigma\rangle\equiv P_\sigma|I\rangle$ for $\sigma\in\mathcal{S}_N$, where $|I\rangle \equiv|\phi_1\rangle\otimes\ldots\otimes|\phi_N\rangle$ for  some arbitrary orthonormal vectors  $|\phi_k\rangle\in\mathcal{H}$, $ \langle \phi_k|\phi_l\rangle =\delta_{k,l}$ (note that  $\langle\sigma|\tau\rangle = \delta_{\sigma,\tau}$ and  $P_\pi = \sum_\tau |\pi\tau\rangle\langle\tau|$ when restricted to   $\mathcal{L}$). Using that  $P_\sigma|\tau\rangle = |\sigma\tau\rangle$ and $\langle \pi|P_\sigma|\pi\rangle = \delta_{\sigma,I}$,  by starting from a trivial identity we get  
\begin{eqnarray}
\label{A15}
J(\sigma) &=& \frac{1}{N!} \sum_{\pi\in \S_N} \langle \pi| \Bigl[\sum_{\tau}J(\tau)P^\dag_\tau \Bigr]P_\sigma |\pi\rangle\nonumber\\
&=&\mathrm{Tr}\left\{ \rho^{(int)} P_\sigma\right\},
\end{eqnarray}
where the  trace is taken  in $\mathcal{H}^{\otimes N} $ and  we have introduced a p.s.d. Hermitian operator (density matrix)  $\rho^{(int)}$ in the Hilbert space $\mathcal{H}^{\otimes N}$    
\be
\rho^{(int)} \equiv \frac{1}{N!}\sum_{\tau\in\S_N} J(\tau)\sum_{\pi\in\S_N} |\pi\rangle\langle\tau\pi|. 
\en{A16}
 Obviously,  $\mathrm{Tr}\{\rho^{(int)}\}=J(I)=1$. Positivity of $\rho^{(int)}$ would follow from the explicit  form (now for bosons and fermions) 
\begin{eqnarray}
\label{A17}
&& \rho^{(int)} = \frac{1}{N!}\sum_{\tau\in\S_N} |\Phi_\tau\rangle\langle\Phi_\tau|,\nonumber\\
&&    |\Phi_\tau\rangle\equiv \sum_{\sigma\in\S_N}\xi(\sigma\tau)P_\sigma\{ |\phi_1\rangle\otimes\ldots\otimes|\phi_N\rangle\},
\end{eqnarray}
where $\xi(\sigma)\equiv\vare(\sigma)\theta^*(\sigma)$ for the factorizing function  $\theta(\sigma)$ (see  Eq. (\ref{AfactJ}) below). 

Let us sketch the prove of factorization of a p.s.d. Hermitian function $J(\sigma)$ (see Ref. \cite{MultLinAlg}). Consider an operator  $\mathcal{J}$ in $\mathcal{L}$   
\be
\mathcal{J} \equiv \sum_{\sigma\in \S_N} J(\sigma)P^\dag_\sigma,
\en{B1}
given by following matrix
\be
  \mathcal{J}_{\nu,\tau} = \langle \nu|\mathcal{J}|\tau\rangle = J(\nu^{-1}\tau),
\en{B2}
which, by assumption that  $J(\sigma)$ is a p.s.d. Hermitian function,  is a p.s.d. Hermitian matrix. Since  operators  having the form given by Eq. (\ref{B1})  constitute an  sub-algebra of operators in $\mathcal{L}$, the p.s.d. Hermitian operator  $\mathcal{J}$ (\ref{B1})  can   be   factorized  by an operator $\mathcal{B}\in \mathcal{L}$   
\be
\mathcal{J} = \mathcal{B}^\dag\mathcal{B}, \quad  \mathcal{B} = \sum_\sigma \theta(\sigma)P^\dag_\sigma.
\en{B4}
By the group   rule $P_\sigma P_\tau = P_{\sigma\tau}$ Eq. (\ref{B4}), in the matrix form, is equivalent to the factorization
\be
J(\sigma) = \sum_{\tau\in\S_N} \theta^*(\tau)\theta(\tau\sigma), \quad \sum_{\sigma\in\S_N} |\theta(\sigma)|^2 = 1.
\en{AfactJ}
The symmetry (\ref{A5})  means that the  factorizing function can be chosen to satisfy  
\be
\theta(\sigma\pi)  = \theta(\sigma), \quad \forall \pi \in \mathcal{S}_{\mathbf{n}}.
\en{ A8}
 (In contrast,  $\theta(\tau\sigma)$ for all $ \tau\in \mathcal{S}_N$ is   the same  factorization with  a different order of  terms).

\section{Derivation of the average probability formulae} 

Though the average quantum detection probabilities  follow from a simple symmetry argument for a Haar-random unitary $U$, the classical case  requires a bit more of insight. Here these results are derived by direct evaluation  demonstrating also  the validity of the classical formula for  $M\gg 1$ (arbitrary $\K$ and $N$), as observed in numerical simulations. The following identity will be employed
\begin{eqnarray}
\label{A18}
&& \langle \prod_{\alpha=1}^N U_{k_\alpha,l_\alpha} U^*_{k^\prime_\alpha,l^\prime_\alpha} \rangle \nonumber\\
&&=  \sum_{\nu,\tau\in\S_N}\W(M,\nu\tau^{-1})\prod_{\alpha=1}^N\delta_{k^\prime_\alpha,k_{\nu(\alpha)}}\delta_{l^\prime_\alpha,l_{\tau(\alpha)}},
\end{eqnarray} 
 where $\W(M,\sigma)$ is the Weingarten function of the unitary group \cite{W,W2} which depends only on the cycle structure of the relative permutation $\sigma=\nu\tau^{-1}$, i.e., the sequence of numbers   $ (c_1(\sigma),\ldots,c_N(\sigma))$ of cycles of lengths $(1,\ldots,N)$ in its  cycle decomposition  (for more details see Ref. \cite{Stanleyap}). By application of Eq. (\ref{A18})  we have 
 \be
  \langle \prod_{\alpha=1}^N U_{k_\alpha,l_\alpha} U^*_{k_{\sigma(\alpha)},l_\alpha} \rangle = \sum_{\nu\in\S_\mathbf{n}}\sum_{\tau\in\S_\mathbf{m}} \W(M,\nu\sigma\tau)
 \en{A20}
 with summation over permutation invariance subgroups $\S_\mathbf{n}$ and $\S_{\mathbf{m}}$ of the input and output indices, respectively. Then, from Eq. (\ref{A4}) we have  for the average over Haar-random network (here the output indices  vary between $1$ and $\K$ and, since we sum over the output modes $l_1,\ldots,l_N$, one of permutations in $J$ is redundant, giving a $N!$ factor)
 \begin{eqnarray}
 \label{A19}
&& \!\!\!\langle p_N(J)\rangle =\frac{N!}{\mu(\mathbf{n})}\!{\sum_{\mathbf{m}}}^\prime\frac{1}{\mu(\mathbf{m})} \!\sum_{\sigma\in\S_N}J(\sigma) \langle \prod_{\alpha=1}^N U_{k_\alpha,l_\alpha} U^*_{k_{\sigma(\alpha)},l_\alpha} \rangle \nonumber\\
 &&= \sum_{\sigma\in\S_N}J(\sigma){\sum_{\mathbf{m}}}^\prime \frac{N!}{\mu(\mathbf{m})}\sum_{\tau\in\S_\mathbf{m}} \W(M,\sigma\tau),
 \end{eqnarray}
where $\sum^\prime$ is over all occupations $\mathbf{m}$ of $\K$ output modes and we have used the property of $J(\sigma)$ in Eq. (\ref{A5}) and  that the number of $\nu\in\S_\mathbf{n}$ is $ |\S_{\mathbf{n}}|=\mu(\mathbf{n})$. Let us now consider separately bosons, fermions and distinguishable particles for a general input $\mu(\mathbf{n})\ge 1$. 

In case of the completely indistinguishable bosons, $J(\sigma) = 1$, we obtain from Eqs. (\ref{A19}) and (\ref{A20}) 
\begin{eqnarray}
\label{A21}
&&\langle p^{(B)}_N\rangle = {\sum_{\mathbf{m}}}^\prime \frac{N!}{\mu(\mathbf{m})}\sum_{\sigma\in\S_N}\sum_{\tau\in\S_\mathbf{m}} \W(M,\sigma\tau) \nonumber\\
&& = \frac{(\K+N-1)!}{(\K-1)!}\sum_{\sigma\in \S_N}\W(M,\sigma),
\end{eqnarray}
 where we have used that $\S_{\mathbf{m}}\subset \S_N$ and $|\S_{\mathbf{m}}| = \mu(\mathbf{m})$ (the first sum  gives  the number of Fock states of $N$ bosons in $\K$ modes scaled  by $N!$). Observing  that for $\K=M$ the probability must be equal to $1$, we get the sum of $\mathcal{W}$-functions     in Eq. (\ref{A21})
\be
\sum_{\sigma\in \S_N}\W(M,\sigma) = \frac{(M-1)!}{(M+N-1)!}.
\en{A22}
Combinig  Eqs.  (\ref{A21}) and (\ref{A22}) we get  the final expression for $\langle p^{(B)}_N\rangle$. 

In case of the completely indistinguishable fermions $J(\sigma) = \mathrm{sgn}(\sigma)$ and $\mu(\mathbf{n}) = \mu(\mathbf{m}) = 1$ (no two or more particles per mode). We have  from Eqs. (\ref{A19}) and (\ref{A20}) 
\begin{eqnarray}
\label{A23}
&&\langle p^{(F)}_N\rangle =    N!{\sum_{\mathbf{m}}}^\prime\sum_{\sigma\in\S_N}\mathrm{sgn}(\sigma) \W(M,\sigma),\nonumber\\
&&=\frac{\K!}{(\K-N)!}\sum_{\sigma\in \S_N} \mathrm{sgn}(\sigma)\W(M,\sigma)
\end{eqnarray}
(the first sum is the number of Fock states of $N$ fermions in  $\K$ modes). Setting    $\K=M$ in Eq. (\ref{A23})   we obtain
\be
\sum_{\sigma\in \S_N} \mathrm{sgn}(\sigma)\W(M,\sigma) = \frac{(M-N)!}{M!}. 
\en{A24} 
Eqs. (\ref{A23}) and (\ref{A24}) result in the final expression for $\langle p^{(F)}_N\rangle$. 

In case of distinguishable   particles from Eqs.  (\ref{A6}) and (\ref{A19}) we get
\begin{eqnarray}
\label{A25}
&&\langle p^{(d)}_N\rangle = \sum_{l_1=1}^\K\ldots\sum_{l_N=1}^\K \sum_{\nu\in\S_\mathbf{n}}\sum_{\tau\in\S_\mathbf{m}}\W(M,\nu\tau)\nonumber\\
&& = \sum_{\nu\in\S_\mathbf{n}}\sum_{\sigma\in\S_N}\W(M,\nu\sigma)\sum_{l_1=1}^\K\dots\sum_{l_N=1}^\K \prod_{\alpha=1}^N\delta_{l_\alpha,l_{\sigma(\alpha)}}\nonumber\\
&& = \sum_{\nu\in\S_\mathbf{n}}\sum_{\sigma\in\S_N}\W(M,\nu\sigma) \K^{\# \sigma}
\end{eqnarray}
where we have set $\# \sigma \equiv c_1(\sigma)+\ldots+c_N(\sigma)$ (the total number of cycles in the cycle  decomposition  of $\sigma$) and  used that $|\S_\mathbf{n}| = \mu(\mathbf{n})$ and  $\prod_{\alpha=1}^N\delta_{l_\alpha,l_{\sigma(\alpha)}} = \sum_{\tau\in\S_\mathbf{m}}\delta_{\sigma,\tau}$. 
 Eq. (\ref{A25}) must  coincide with Eq. (\ref{A21}) for all particles in the same input mode, $\mu(\mathbf{n}) =N!$, since  the limit of distinguishable particles Eq. (\ref{A6}) is obtained by making  identical   particles from different modes distinguishable, while  particles from the  same input mode behave as  the completely indistinguishable bosons. On the other hand, for a single-particle input  it has a form quite different from that of Eq. (\ref{A21}) for a similar  input of the completely indistinguishable bosons.   Apparently, the general  expression would have quite a  cumbersome form.  Consider the special case of single particles.  Using an asymptotic form  of  $\W$  for $M\gg 1$   \cite{W2} 
 \begin{eqnarray}
 \label{A26}
  \W(M,\sigma) &= &\frac{(-1)^N}{M^{2N}}\prod_{s=1}^N(-Mg_s)^{c_s(\sigma)}\left(1+ O\left(\frac{1}{M^2}\right)\right)\nonumber\\ 
  g_s &=& \frac{(2s-2)!}{s!(s-1)!},
 \end{eqnarray}
 we obtain from Eq. (\ref{A25}) the leading term in the average classical probability as a cycle sum 
 \be
 \langle p^{(d)}_N\rangle \approx \frac{(-1)^N}{M^{2N}}\sum_{\sigma\in\S_N}\prod_{s=1}^N(-\K{}Mg_s)^{c_s(\sigma)} = \frac{(-1)^NN!}{M^{2N}}Z_N,
  \en{A27} 
with ($t_s = -\K{}Mg_s$)
\be
Z_N \equiv \frac{1}{N!} \sum_{\sigma\in\S_N} \prod_{s=1}^N t_s^{c_s(\sigma)}.
\en{A28} 
The cycle sum is evaluated by the generating function method  (see, for instance,  Ref. \cite{Stanleyap}) which satisfies the following   identity
\be
\mathcal{F}(x) \equiv \sum_{N\ge 1} Z_Nx^N = \exp\left\{ \sum_{s\ge1}t_s\frac{x^s}{s}\right\}. 
\en{A29}
In our case  (after evaluation of a table sum \cite{PBM})
 we get  $\mathcal{F}(x)$ to be  
\begin{eqnarray}
\label{A30}
&&\mathcal{F}(x) = \exp\left\{-\K M \sum_{s=1}^\infty \frac{(2s-2)!}{(s!)^2}x^s \right\} \\
&& = \left(\frac{2}{1+\sqrt{1-4x}}\exp\left\{ -[1-\sqrt{1-4x}]\right\} \right)^{\K{}M}.\nonumber
\end{eqnarray} 
For $\K M\gg 1$,   the leading order  reads
\be
Z_N = \frac{1}{N!}\frac{d^N \mathcal{F}(0)}{dx^N}\approx \frac{ (-\K{}M)^N}{N!},
\en{A31}
due to the fact that    the derivative of $\mathcal{F}(x)$, 
\be
\frac{d\mathcal{F}(x)}{d x} =  \frac{-2\K{}M}{\sqrt{1-4x}(1+\sqrt{1-4x})}\mathcal{F}(x),
\en{A32}
has   as a factor a slowly varying function of $|x|\ll 1$  (in comparison with $\mathcal{F}(x)$). Hence \mbox{$\frac{d^N\mathcal{F}(0)}{d x^N} = (-\K M)^N\left[ 1 - N(N-1)/2\K M + \ldots\right]$.} Substituting Eq. (\ref{A31}) into Eq. (\ref{A27}) we obtain  $\langle p^{(d)}_N\rangle $ up to a factor  $(1+O(\frac{N^2}{\K M}))$.

\section{Classical computations   for  the assessment protocol are only polynomial in $N$}

The assessment protocol for the standard version of BS  is dependent on $p^{(B)}_N = \mathrm{per}(H)$, where $H$ is a p.s.d. Hermitian matrix built from  a  submatrix  of a network matrix $U$ according to Eq. (6) of the main text. Such a matrix $H$ can  be rewritten as 
 \be
 H_{\alpha,\beta} = \delta_{\alpha,\beta} -  \sum_{l=\K+1}^{M} U_{k_\alpha,l}U^*_{k_\beta,l}\equiv (I - \Phi)_{\alpha,\beta},
 \en{P1}
where   $k_1,\ldots,k_N$ are  input indices.    Below we show that, in the dilute limit, $NL/M\to 0$ as $N\to \infty$, with a high probability over a choice of $L=M-\K$ output modes,  only polynomial in $N$ computations are required to estimate $\mathrm{per}(H)$ of Eq. (\ref{P1}) to an arbitrary error $\epsilon$ polynomial in $N^{-1}$~\footnote{Since  only polynomial number of runs of a BS device can be employed in an efficient assessment protocol, only a polynomial in $N^{-1}$ accuracy  in  probability $p^{(B)}_N$ can be achieved in an experiment.}.   Since  we can select    $L$ output modes completely arbitrary, one can  always find  a suitable  set  in a polynomial number of trials. Moreover, an approximation formula to the required polynomial accuracy is given (Eq. (\ref{P8}) below).   

 We will use   a general formula for the matrix permanent of a sum of two matrices \cite{Minc}, in our case
\be
\mathrm{per}(H) = 1 + \sum_{r=1}^N(-1)^r \sum_{\alpha_1<...<\alpha_r} \mathrm{per}(\Phi[\vec{\alpha},\vec{\alpha}]),
\en{P2}
where $\Phi[\vec{\alpha},\vec{\alpha}]$ is an $r$-dimensional  submatrix of $\Phi$ built on the rows and columns $\vec{\alpha} = (\alpha_1,\ldots,\alpha_r)$. Now, since for $NL/M\ll 1$ the result on the r.h.s. of Eq. (\ref{P2}) is close to $1$ (see Eq. (12) of the main text), we expect that only a few terms in this expression are required for estimating $\mathrm{per}(H) $ to a polynomial in $N^{-1}$  error.

Indeed, let us consider how many terms on the r.h.s. of Eq. (\ref{P2}) are required on average over the Haar-random networks. By application of the general result given by Eq. (12) in the main text (see also the previous section) we have 
\be
\langle \mathrm{per}(\Phi[\vec{\alpha},\vec{\alpha}]) \rangle =  \frac{L(L+1)\cdot\ldots\cdot(L+r-1)}{M(M+1)\cdot\ldots\cdot(M+r-1)},
\en{P3}
hence, for $r\ll \sqrt{N}$ and $r\ll \sqrt{L}$ we obtain 
\be
\langle T_r\rangle \equiv \sum_{\alpha_1<...<\alpha_r}\langle \mathrm{per}(\Phi[\vec{\alpha},\vec{\alpha}]) \rangle = (1+\varepsilon)\frac{1}{r!} \left(\frac{NL}{M}\right)^r,
\en{P4}
 where $\varepsilon \sim r^2[1/N + 1/L]\ll 1$. 
 
To estimate $\mathrm{per}(H) $ to an error $\epsilon = O(N^{-\kappa})$, for  $\kappa>0$,   we can truncate the sum on the r.h.s. of Eq. (\ref{P2}) at  an order $s$ satisfying $ \langle T_s\rangle \ll \epsilon$.  To have a single scale $N^{-1}$, let us  concentrate on the case  given by $M=b N^{2+\delta}$ for $b=O(1)$ and $\delta>0$ and $L = aN$  with  $a = O(1)$. In this case, averaging  over Haar-random $U$  gives  $\langle\mathrm{per}(H) \rangle=   1-O(N^{-\delta})$. The truncation order  $s$ must satisfy
\be
\langle T_s\rangle \approx \frac{1}{s!}\left(\frac{a}{bN^\delta}\right)^s \ll   \epsilon =  O(N^{-\kappa}),
\en{P6} 
i.e., for $N\to \infty$ one must choose $s > {\kappa}/{\delta}$. Let us set $s = (\kappa+1)/\delta$ (note that $s$ is \textit{an independent of $N$ number}) then for $N\gg 1$  the l.h.s. of Eq. (\ref{P6}) is smaller by a factor $O(N^{-1})$ than the r.h.s..   
Now, let us show that  the number of required computations (flops) $\mathcal{C}_N$   in the summation of the r.h.s. of Eq. (\ref{P2})  to the above truncation order $s = O(1)$ scales only polynomially in $N$ as $N\to \infty$.  Since computation of a matrix permanent of dimension $r$ requires $r2^r$ operations  (flops) by Ryser's algorithm  \cite{Minc}, we can estimate    the  total  number of flops  as follows. First we note that  Eq. (\ref{P2}) truncated at an order $s$   has an equivalent simpler expression (to the same accuracy)
\be
\mathrm{per}(H) \approx 1 +   \sum_{\alpha_1<...<\alpha_s} \left\{\mathrm{per}(\Phi[\vec{\alpha},\vec{\alpha}]) -1\right\}
\en{P8}
(we have taken into account that   indices \mbox{ $\alpha_1<...<\alpha_r$} with $r\le s$ are contained in one of \mbox{$\alpha_1<...<\alpha_s$} and used Eq. (\ref{P2})   to  sum the terms for each subset  \mbox{$\alpha_1<...<\alpha_s$}). From Eq. (\ref{P8}) we obtain  
 \be
\mathcal{C}_N   \sim \frac{N!}{(N-s)!s!}s2^s  = O\left(N^\frac{\kappa+1}{\delta}\right).
 \en{P7}
where we have used $s = (\kappa+1)/\delta$ for the approximation error $\epsilon  = O(N^{-\kappa})$.

The estimates in Eqs. (\ref{P7}) and (\ref{P8}) (and that of Eq. (\ref{P4})) have a high probability, since by  Chebyshev's inequality (for $N\gg 1$)
 \be
 Pr(T_s>\epsilon)\le \frac{\langle T_s\rangle}{\epsilon}= O(N^{-1})\ll 1,
 \en{Cheb}
 where we have used Eq.~(\ref{P6})~\footnote{Generally, given a p.s.d. Hermitian matrix $A$, representable in  the form $A =  U\Lambda U^\dag$  with $U$ being a $M$-dimensional unitary matrix and $\Lambda = \mathrm{diag}(1,\ldots,1,0,\ldots,0)$, with $\K$ ones, where  $M= O(N^{2+\delta})$ for some $\delta>0$ and  $M-\K = O(N)$,   and  a preset  probability of success scaling polynomially in $N^{-1}$,   the  described algorithm  returns  $(1+O(\epsilon))\mathrm{per}(H)$ with the preset probability for any $N$-dimensional principal submatrix $H$ of $A$  in a number of flops scaling polynomially in both $N$  and  $1/\epsilon$.}.  Since   we  select $L=M-\K$ output modes for the protocol  completely arbitrary from the  output modes of a network,  if it turns out that for our  choice of $L$ modes for a particular network $U$ we  need more than $\mathcal{C}_N$ flops, we can  select again (e.g., by using the Gaussian approximation \cite{AAap}, the probability that  no suitable choice  can be found among  $O(N)$  non-intersecting subsets of size $L=O(N)$  from the total of $M = O(N^{2+\delta})$   modes decreases to zero  at  least as $O(N^{-N})$).  
 
One important example is the test against the distinguishable  particles, for which  one can set $\kappa = 2(1+\delta)$.   Indeed, the difference $\langle p^{(B)}_N-p^{(d)}_N\rangle $ between the full quantum and the  classical cases  (see the main text)  reads in our case   
\be
\langle p^{(B)}_N-p^{(d)}_N\rangle  =    O\left(\frac{LN^2}{\K M}\right) = O\left(N^{-1-2\delta}\right),
\en{P9}
therefore, to  distinguish the two cases requires using   only polynomial  computations by the above analysis. 

\section{Effect of the distinguishability error  on  probability $p_N$  and  its Haar-average  }
\label{sec6}

Consider an input state consisting of single bosons from $N$    sources (we will consider uncorrelated sources, the case of classically correlated ones is a trivial extension). The corresponding internal state  reads  
\be
\rho^{(int)}  =  \rho^{(1)}\otimes\ldots\otimes \rho^{(N)}, 
\en{S1} 
where $\rho^{(\alpha)}$ is an internal state of the  $\alpha$th boson. The ideal bosonic input corresponds (in the case of only classically correlated inputs) to all $\rho^{(\alpha)} = |\phi\rangle\langle\phi|$. Hence, let us assume that each source produces a boson in a state  close to a pure state  
\be
\rho^{(\alpha)} = |\phi\rangle\langle\phi| - \delta\rho^{(\alpha)}
\en{S2}
with the indistinguishability fidelity   defined as ${F}_\alpha \equiv  \langle \phi|\rho^{(\alpha)}|\phi\rangle$. To derive the scaling law for small variations 
$1-F_\alpha = \langle\phi|\delta\rho^{(\alpha)}|\phi\rangle \ll 1$ we will assume that sources emit  bosons with  a mean fidelity $F$. 
Keeping only the first order term in $1-F$ we obtain from Eq. (\ref{S1})
\begin{eqnarray}
\label{S3}
&&\rho^{(int)} \approx \left(|\phi\rangle\langle\phi| \right)^{\otimes N}  \\
&& - \sum_{\alpha=1}^N  \left(|\phi\rangle\langle\phi| \right)^{\otimes (\alpha-1)}\otimes \delta\rho^{(\alpha)}\otimes  \left(|\phi\rangle\langle\phi| \right)^{\otimes (N-\alpha)}. \nonumber
\end{eqnarray}
Taking into account that $\mathrm{Tr}\{\delta \rho^{(\alpha)}\} = 0$,  we obtain from the definition (\ref{AJ}) and Eq. (\ref{S3}) up to the first oder  in $1-F$
\be
J(\sigma)  = \mathrm{Tr}\{\rho^{(int)} P_\sigma\} \approx 1 - (1-F)[N - c_1(\sigma)],
\en{S4}
where $c_1(\sigma)$ is the number of $1$-cycles (fixed points) in the cycle decomposition of permutation $\sigma$ ($c_1(\sigma)$  terms in the sum in Eq. (\ref{S3}) give  zero). 

Let us first derive the scaling law for the standard version of BS. Substitution of Eq. (\ref{S4}) into Eq. (\ref{A7}) gives  (with $p^{(B)}_N=\mathrm{per}(H)$)
\be
p_N(J) \approx p^{(B)}_N - (1-F)\left [Np^{(B)}_N - \sum_{\sigma\in \S_N} c_1(\sigma) \prod_{\alpha=1}^N H_{\alpha,\sigma(\alpha)}\right].
\en{ES1}
The second term  in the brackets on the r.h.s. of Eq. (\ref{ES1}) is different from the plain matrix permanent by a factor at each term, counting how many diagonal elements of matrix  $H$  it contains. It can be represented in a computable form by observing that the same counting is done by multiplying diagonal elements of $H$ by a dummy variable $x$ and  application of a derivative w.r.t. $x$ at $x=1$. Thus we have derived
\be
p_N(J) \approx p^{(B)}_N - (1-F)\left [Np^{(B)}_N - \frac{d}{dx} \mathrm{per}\{H(x)\}_{x=1} \right]
\en{ES2}
with
\be
H(x) \equiv    H + (x-1) \mathrm{diag}(H_{11},\ldots,H_{NN}) 
\en{ES3}
The derivative can be evaluated from the Lagrange representation of the polynomial $\mathrm{per}\{H(x)\}$, thus it requires approximating   $N+1$ matrix permanents of the p.s.d. Hermitian matrix $H(x)$ at distinct values of $x$ to some small error $\epsilon$, which can be chosen inversely  polynomial in $N$. In  the previous section it is shown that  this requires only  polynomial in $N$ computations.

Now, let us derive a variant of the scaling law which is specifically tailored for the scattershot BS. Assuming that $M\gg N^2$ (so that one can use the Gaussian approximation for the elements of the Haar-random unitary matrix $U$ \cite{AAap}) we use  the expression (\ref{S4})  into the average probability of detecting all $N$ photons in $\K$ output modes (i.e., $\mathbf{m} = (m_1,\ldots,m_\K,0,\ldots,0)$, see   also Eqs. (\ref{A4}) and (\ref{A7}))  and recall  the   formula   for the   average probability $ \langle p^{(B)}_N\rangle$ (in the dilute limit $M\gg N^2$):
\begin{eqnarray}
\label{S5}
&&\langle p_N(J) \rangle ={\sum_{\mathbf{m}}}^\prime \frac{N!}{\mu(\mathbf{m})}\sum_{\sigma\in \S_N}J(\sigma)\langle \prod_{\alpha=1}^NU^*_{k_{\sigma(\alpha)},l_ \alpha}U_{k_\alpha,l_\alpha}\rangle\nonumber\\
&& \approx  \langle p^{(B)}_N\rangle  - (1-F)\frac{N!}{M^N}{\sum_{\mathbf{m}}}^\prime\frac{1}{\mu(\mathbf{m})}\sum_{\sigma\in\S_{\mathbf{m}}}[N-c_1(\sigma)]\nonumber\\
&& = \langle p^{(B)}_N\rangle  - (1-F)\frac{N(N-1)}{M} \langle p^{(B)}_{N-1}\rangle
\end{eqnarray}
(note that the final result has an appealing form, apparently indicating that it might be valid without assuming the dilute limit $M\gg N^2$). 
Mathematical details  in the derivation of Eq. (\ref{S5}) are as follows. We have  used an approximate independence of $2N$ matrix elements of $U$ for $M\gg N^2$ and the Gaussian approximation \cite{AAap}  in computing the average of  the product, with  $\langle |U_{kl}|^2\rangle \approx 1/M$ and  for  single bosons at input ($k_\alpha \ne k_\beta$ for $\alpha\ne \beta$)
\be
\langle \prod_{\alpha=1}^NU^*_{k_{\sigma(\alpha)},l_ \alpha}U_{k_\alpha,l_\alpha}\rangle \approx \frac{1}{M^N}\sum_{\pi\in\S_{\mathbf{m}}}\delta_{\sigma,\pi},
\en{S6}  
since  all permutations of the output indices   belonging  to the output subgroup 
$\S_{\mathbf{m}}$ equally contribute to the result. The  double sum in the line before the last in Eq. (\ref{S5}) (over all $\mathbf{m}$ and the  subgroup $S_{\mathbf{m}}$) 
\be
R = {\sum_{\mathbf{m}}}^\prime\frac{1}{\mu(\mathbf{m})}\sum_{\sigma\in\S_{\mathbf{m}}}[N-c_1(\sigma)]
\en{S7}
is computed as follows.  We have a sum over permutations 
\begin{eqnarray}
\label{S8}
&&\frac{1}{\mu(\mathbf{m})}\sum_{\sigma\in\S_{\mathbf{m}}}c_1(\sigma) = \frac{1}{\mu(\mathbf{m})}\sum_{\sigma_1\in\S_{m_1}}\ldots\sum_{\sigma_\K\in\S_{m_\K}}\sum_{l=1}^\K c_1(\sigma_l)\nonumber\\
&&= \K - \sum_{l=1}^\K \delta_{m_l,0},
\end{eqnarray}
i.e., the average number of fixed points  (over all permutations)  in  each  permutation group $\S_{m_l}$ is equal to $1$ \cite{Stanleyap}, whereas if $m_l=0$ there is no corresponding contribution. The first sum in Eq. (\ref{S7}) is then easily computed 
\begin{eqnarray}
\label{S10}
R &=& {\sum_{\mathbf{m}}}^\prime\left(N -\K + \sum_{l=1}^\K \delta_{m_l,0}\right) \nonumber\\
&=& (N-\K) \frac{(\K+N-1)!}{N!(\K-1)!} + \K \frac{(\K+N-2)!}{N!(\K-2)!}\nonumber\\
&=& (N-1)\frac{(\K+N-2)!}{(N-1)!(\K-1)!},
\end{eqnarray}
where the summation gives the number of Fock  states of $N$  bosonic particles distributed over, respectively, $\K$ and $\K-1$ modes. 

\section{Equivalent description of a lossy linear network}
 
Below we focus on bosonic particles, for fermions one should replace the commutators below by the anti-commutators.  A realistic network  $U$ is non-unitary due to (generally path-dependent)  losses of particles,  its action is described not just by   input and output mode operators, $a_1,\ldots, a_M$ and $b_1,\ldots,b_M$, but also by some additional operators $f_1,\ldots, f_M$ accounting for losses:
\be
a^\dag_k = \sum_{l=1}^M U_{k,l}b^\dag_l + f^\dag_k,
\en{Anet}
where operators $f_k$ and $f^\dag_k$    commute   with the creation and annihilation operators corresponding to network modes: $[f_k,a_l] = [f_k,b_l] = [f^\dag_k,a_l] = [f^\dag_k,b_l]=0$ \cite{FCR}.  Using the latter we obtain
\be
[f_k,f_j]=0,\quad [f_k,f^\dag_j] = \delta_{k,j} - \sum_{l=1}^M U^*_{k,l}U_{j,l}.
\en{Comf}
Eq. (\ref{Comf}) can be satisfied if we expand the loss operators $f^\dag_1,\ldots,f^\dag_M$ over some additional creation operators
\be
f^\dag_k = \sum_{l=1}^M V_{k,l}b^\dag_{M+l}, 
\en{Expf} 
where $V$ is a   matrix satisfying the following matrix  equation (valid both for bosons and  fermions)
\be
VV^\dag = I - UU^\dag,
\en{V}
where $I$ is the unit matrix and  $U^\dag$ denotes the Hermitian conjugate to matrix $U$. Notice that Eq. (\ref{V}) requires  that the singular values of $U$ be bounded by $1$, which is the necessary   and sufficient condition for an arbitrary complex matrix $U$ to describe a passive linear quantum network.  There is a  polar decomposition, $U= \sqrt{A}\mathcal{U}$, where  $\mathcal{U}$ is a unitary matrix and  $A = UU^\dag$  a Hermitian matrix   (describing  losses in the network) with the eigenvalues bounded by $1$.  

The expansion in Eq. (\ref{Expf})   means that one can imbed an arbitrary  (non-unitary) linear $M$-mode network into a $2M$-mode unitary one~\footnote{Moreover,  analysis of necessary conditions on such embedding shows that one cannot reduce the size of an embedding network for non-singular  matrices $U$.}.    The following  embedding   unitary network seems to be the simplest one 
\be
\hat{U} =  \left(\begin{array}{cc}U& V\\
-V^\dag \mathcal{U} & D  \end{array}\right),  
\en{EmbU}
here the diagonal matrix  $D = \mathrm{diag}(\eta_1,\ldots,\eta_M)$, \mbox{$0\le \eta_k\le 1$,} is composed of the  square-roots of singular values of $U$ (eigenvalues of $\sqrt{A}$, since  $A = UU^\dag$),     $V = SQ$,  with the  unitary matrix $S$ containing  the eigenvectors of $\sqrt{A}$, $\sqrt{A} = S D S^\dag$,  $Q = \mathrm{diag}(\sqrt{1-\eta_1^2},\ldots,\sqrt{1-\eta_M^2})$, and $\mathcal{U}$ is from the polar decomposition $U = \sqrt{A}\mathcal{U}$. Matrix $\hat{U}$ can be also rewritten in a product form
\be
\hat{U} = \left(\begin{array}{cc}S & 0\\
0& I  \end{array}\right) \left(\begin{array}{cc}D &Q\\
-Q& D  \end{array}\right) \left(\begin{array}{cc}S^\dag & 0\\
0& I  \end{array}\right) \left(\begin{array}{cc}\mathcal{U} & 0\\
0& I  \end{array}\right),
\en{EmbU1}
from  which it is evident that $\hat{U}$ is  unitary. Note that in description of a lossy network $U$, the unitary matrix $\hat{U}$ has vacuum input in the modes $\{M+1,\ldots,2M\}$ and output modes $\{M+1,\ldots,2M\}$ are not accessible ``loss channels".  The embedding matrix of Eq. (\ref{EmbU}) reduces to a matrix appeared in Ref. \cite{SLoss} in the  special case of path-independent losses, i.e., a diagonal loss matrix $A$.

\end{document}